\documentclass[twocolumn,showpacs,preprintnumbers,amsmath,amssymb]{revtex4}
\usepackage{amsmath}

\usepackage{graphicx}
\usepackage{dcolumn}
\usepackage{bm}

\begin{document}

\def\vsigma{{\hbox{\boldmath $\sigma$}}}
\def\vlambda{{\hbox{\boldmath $\lambda$}}}

\title{Two-pion interferometry for the granular sources in ultrarelativistic heavy ion collisions}

\author{Wei-Ning Zhang$^{1,2}$\footnote{wnzhang@dlut.edu.cn}}
\author{Hong-Jie Yin$^1$}
\author{Yan-Yu Ren$^2$}

\affiliation{$^1$School of Physics and Optoelectronic Technology,
Dalian University of Technology, Dalian, Liaoning 116024, China\\
$^2$Department of Physics, Harbin Institute of Technology, Harbin,
Heilongjiang 150006, China\\
}

\begin{abstract}

We investigate the two-pion interferometry in ultrarelativistic
heavy ion collisions in the granular source model of quark-gluon
plasma droplets.  The pion transverse momentum spectra and HBT radii
of the granular sources agree well with the experimental data of the
$\sqrt{s_{NN}}=200$ GeV Au-Au and $\sqrt{s_{NN}}=2.76$ TeV Pb-Pb
most central collisions.  In the granular source model the larger
initial system breakup time may lead to the larger HBT radii $R_{\rm
out}$, $R_{\rm side}$, and $R_{\rm long}$.  However, the large
droplet transverse expansion and limited average relative emitting
time of particles in the granular source lead to small ratios of the
transverse HBT radii $R_{\rm out}/R_{\rm side}$.

\end{abstract}


\maketitle

Hanbury-Brown-Twiss (HBT) interferometry is a useful tool to probe
the space-time geometry of the particle-emitting sources in high
energy heavy ion collisions \cite{CYW94,UAW99,RMW00,MAL05}. The
experimental results of the HBT measurements for the Au-Au
collisions at the high energies of the Relativistic Heavy Ion
Collider (RHIC) indicate that it is hard to describe the source
space-time by a simple evolution model
\cite{STA01a,PHE02a,PHE04a,STA05a}.  HBT interferometry data provide
strong constraints for the models of source space-time.  Recently,
the HBT measurement for the $\sqrt{s_{NN}}=2.76$ TeV Pb-Pb most
central collisions at the Large Hadron Collider (LHC) is performed
\cite{ALI11}.  A consistent explanation to the HBT data of the LHC
and RHIC experiments is required naturally for the source models,
which will be helpful to understand the initial condition, source
evolution, and particle freeze-out in ultrarelativistic heavy ion
collisions.

In Refs. \cite{WNZ06,WNZ09}, the granular source model of
quark-gluon plasma (QGP) droplets \cite{WNZ04} is developed to
explain the RHIC HBT data \cite{PHE04a,STA05a}.  In this work we
investigate the two-pion HBT interferometry in ultrarelativistic
heavy ion collisions in the granular source model of QGP droplets.
Our results indicate that the granular source for the LHC Pb-Pb
collisions may have the same initial droplet temperature and
velocity formula as those for the RHIC Au-Au collisions, but a
larger initial system breakup time.  The consistent granular source
model reproduces the pion transverse momentum spectra and HBT radii
in the most central collisions of the RHIC \cite{PHE04,STA04,STA05a}
and LHC \cite{ALI11a,ALI11} experiments.

In ultrarelativistic heavy ion collisions, the system at central
rapidity may reach a local equilibrium at a very short time
$\tau_0$, then fast expand in the beam direction ($z$-axis). Because
of the initial fluctuation, the local-equilibrium system is not
uniform in space \cite{HJD02,OSJ04}.  It may form many tubes along
the beam direction during the fast longitudinal expansion, and
finally fragment into many droplets (see Fig.1 of Ref. \cite{WNZ06})
due to the ``sausage" instability and surface tension \cite{WNZ06}.
On the other hand, the rapidly increased bulk viscosity in the QGP
near the phase transition may also leads to the system breakup
\cite{GTO08}.

We assume that the system fragments and forms a granular source of
many QGP droplets at a time $t_0 (>\tau_0)$.  On the basis of the
Bjorken hypothesis \cite{JDB83}, the longitudinal velocity and
rapidity of the droplets at $t_0$ are
\begin{equation}
\label{vdropz} v_{dz}=z_0/t_0,~~~~\eta_0=\frac{1}{2} \log \frac{t_0+z_0}{t_0 -z_0}\,,
\end{equation}
and the transverse velocity of the droplets may be expressed as
\cite{GBA83,WNZ09}
\begin{equation}
\label{vdropt} v_{d \perp}=a_T \bigg(\frac{\rho_0}{{\cal R}_{\perp 0}}\bigg)^{b_T} \sqrt{1-v_{dz}^2}\,,
\end{equation}
where $z_0$ and $\rho_0$ are the longitudinal and transverse
coordinates of the droplet at the breakup time $t_0$,  ${\cal
R}_{\perp 0}$ is the maximum transverse radius of the system at
$t_0$.  In Eq. (\ref{vdropt}), the quantities $a_T$ and $b_T$ are
the magnitude and power parameters of the transverse velocity which
will be determined by the data of particle transverse momentum
spectra.

In our model calculations, the initial droplet radius in droplet
local frame satisfies a Gaussian distribution with standard
deviation $\sigma_d$, and the initial droplet centers are assumed
distributing within a cylinder along the beam direction by
\cite{WNZ06,WNZ09}
\begin{eqnarray}
\label{dNr}
\frac{dN}{2\pi\rho_0\,d\rho_0} \propto
\left[1-\exp\,(-\rho_0^2/\Delta{\cal
R}_{\perp}^2)\right]\theta({\cal R}_{\perp}-\rho_0)\,,
\end{eqnarray}
where ${\cal R}_{\perp}$ and $\Delta{\cal R}_{\perp}$ are the
initial transverse radius and shell parameter of the granular source
\cite{WNZ06,WNZ09}.  Because of the longitudinal boost-invariant in
ultrarelativistic heavy ion collisions, we may obtain the initial
coordinate $z_0$ of the droplet by the longitudinal distribution
\begin{eqnarray}
\label{dNz}
\frac{dN}{dz_0}=\frac{dN}{d\eta_0}\frac{d\eta_0}{dz_0}\propto
1\cdot\frac{t_0}{t_0^2-z_0^2},~~~~|z_0| < \sqrt{t_0^2 -\tau_0^2}.
\end{eqnarray}

The evolution of the granular source after $t_0$ is the
superposition of all the evolutions of the individual droplets, each
of them is described by relativistic hydrodynamics with an initial
local energy density $\epsilon_0$ and the equation of state (EOS) of
the entropy density with a cross over between the QGP and hadronic
gas \cite{JPB87,ELA96,DHR96}. The values of the EOS parameters used
in the calculations are taken as the same as in Ref. \cite{WNZ09}.
In order to include the pions emitted directly at hadronization and
decayed from resonances later, we let the pions freeze-out within a
wide temperature region with the probability \cite{WNZ09}
\begin{eqnarray}
\label{Pt}
\frac{dP_f}{dT} &\propto& f_{\rm dir}\,
e^{-(T_c-T)/\Delta T_{\rm dir}} + (1-f_{\rm dir}) \nonumber \\
&\times& e^{-(T_c-T)/\Delta T_{\rm dec}}\,,~~(T_c > T > 80~{\rm MeV})\,,~~
\end{eqnarray}
where $T_c$ is the transition temperature, $f_{\rm dir}$ is a
fraction parameter for the direct emission, $\Delta T_{\rm dir}$ and
$\Delta T_{\rm dec}$ describe the widths of temperature for the
direct and decayed pion emissions.  They are taken to be $T_c=170$
MeV, $f_{\rm dir}=0.85$, $T_{\rm dir}=10$ MeV, and $T_{\rm dec}=90$
MeV \cite{WNZ09}.

\begin{figure} [h]
\vspace*{3 mm}
\includegraphics[scale=0.55]{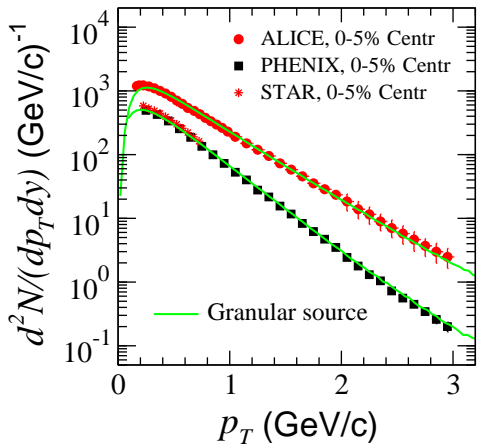}
\vspace*{1mm} \caption{(Color online) The pion transverse momentum
distribution of the granular source and the experimental data of
$\sqrt{s_{NN}}=200$ GeV Au-Au (PHENIX \cite{PHE04} and STAR \cite{STA04})
and $\sqrt{s_{NN}}=2.76$ TeV Pb-Pb (ALICE \cite{ALI11a}) most central
collisions.} \label{Figsp}
\end{figure}

In Fig. \ref{Figsp}, we show the pion transverse momentum spectra
calculated with the granular source model and the experimental data
of $\sqrt{s_{NN}}=200$ GeV Au-Au \cite{PHE04,STA04} and
$\sqrt{s_{NN}}=2.76$ TeV Pb-Pb \cite{ALI11a} most central
collisions.  Assuming that the systems fragment when the local
energy density decreases at a certain value, we take the initial
droplet temperature $T_0=200$ MeV \cite{WNZ09} in the calculations.
The corresponding initial energy density of the droplets,
$\epsilon_0$, is 2.24 GeV/fm$^3$.  The standard deviation $\sigma_d$
for the droplet radius distribution is taken to be 2.5 fm.  For the
granular source in the RHIC collisions, $\tau_0$ and the breakup
time $t_0$ are taken to be 0.8 and 4.3 fm/c, and the source
parameters ${\cal R}_{\perp}$ and $\Delta{\cal R}_{\perp}$ are taken
to be 6.5 and 3.4 fm, respectively.  For the granular source in the
LHC collisions, these parameters are taken to be $\tau_0=0.4$ fm/c,
$t_0=8.0$ fm/c, ${\cal R}_{\perp}=7.8$ fm, and $\Delta{\cal
R}_{\perp}=5.5$ fm.  By comparing the transverse momentum spectra of
the granular sources with the data of $\sqrt{s_{NN}}=200$ GeV Au-Au
\cite{PHE04,STA04} and $\sqrt{s_{NN}}=2.76$ TeV Pb-Pb \cite{ALI11a}
most central collisions, we fix the parameters of the droplet
transverse velocity $a_T=0.65$ and $b_T=1.40$ the same for both the
two energy sources in Eq. (\ref{vdropt}), where ${\cal R}_{\perp 0}$
is taken to be 8.0 fm in the calculations.

\begin{figure} [h]
\vspace*{1mm}
\includegraphics[scale=0.43]{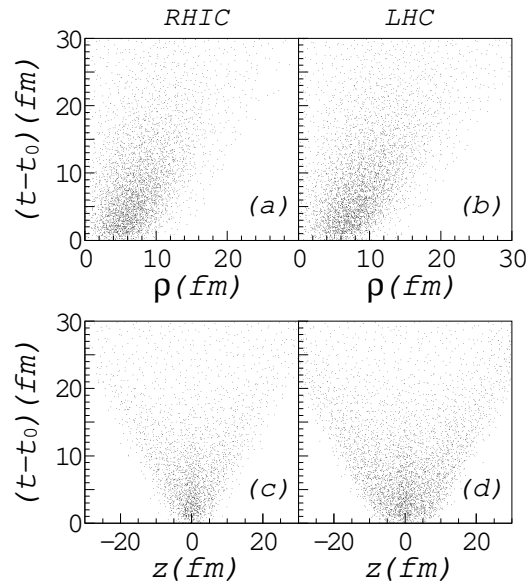}
\vspace*{-5mm} \caption{The space-time distribution of pion-emitting
points in the granular source model.} \label{Figdenstrz}
\end{figure}

Figure \ref{Figdenstrz} shows the space-time distributions of the
pion-emitting points in the granular source model.  Because of the
higher collision energy, the granular source for the LHC collisions
forms (system energy density decreases to a certain value) at the
larger $t_0$, and hence has a wider $z_0$ distribution (see Eq.
(\ref{dNz})) and larger ${\cal R}_{\perp}$ value (wider $\rho_0$
distribution, see Eq. (\ref{dNr})). These lead to the wider
distributions of the longitudinal and transverse source points for
the LHC granular source than those for the RHIC granular source.

\begin{figure} [h]
\vspace*{1mm}
\includegraphics[scale=0.6]{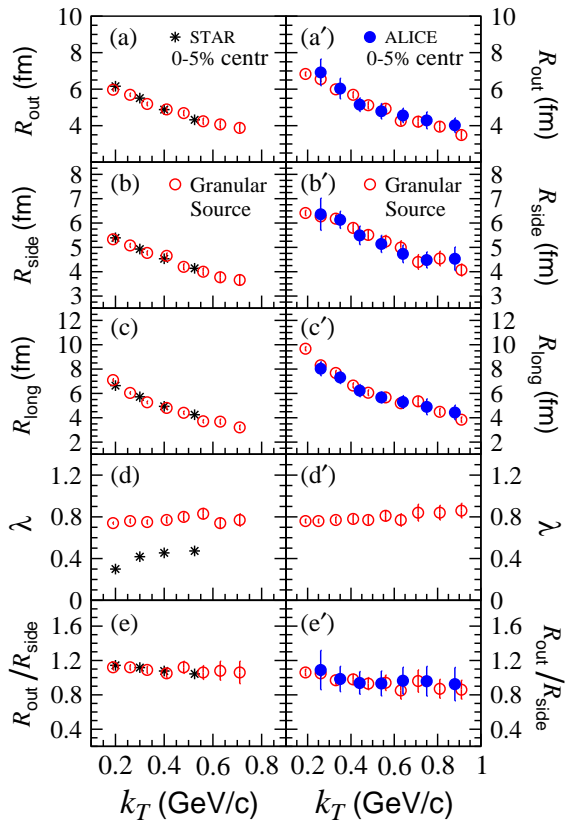}
\vspace*{1mm} \caption{(Color online) (a) The two-pion interferometry
results of the granular sources and the experimental HBT data for the
most central collisions of $\sqrt{s_{NN}}=200$ GeV Au-Au \cite{STA05a}
and $\sqrt{s_{NN}}=2.76$ TeV Pb-Pb \cite{ALI11}.} \label{Fighbtr}
\end{figure}

In Fig. \ref{Fighbtr}, we show the HBT radii $R_{\rm out}$, $R_{\rm
side}$ and $R_{\rm long}$ in the ``out", ``side", and ``long"
directions \cite{GBE88,SPR90}, and the chaotic parameter $\lambda$
of the granular sources as functions of the transverse pion pair
momentum, $k_T=|{\bf p}_{1T}+{\bf p}_{2T}|/2$.  They are obtained by
fitting the two-pion correlation functions in different $k_T$
regions with the formula
\begin{equation}
C(q_{\rm out}, q_{\rm side}, q_{\rm long})=1+\lambda \,e^{-q^2_{\rm
out} R^2_{\rm out} -q^2_{\rm side} R^2_{\rm side} -q^2_{\rm long}
R^2_{\rm long}},
\end{equation}
in the longitudinal comoving system (LCMS) \cite{MAL05}.  The HBT
data of the most central collisions of $\sqrt{s_{NN}}=200$ GeV Au-Au
(STAR \cite{STA05a}) and $\sqrt{s_{NN}}=2.76$ TeV Pb-Pb (ALICE
\cite{ALI11}) are also shown.  In our calculations, we use the same
cuts as in the experimental analyses \cite{STA05a,ALI11}, that is
the pion rapidity is limited with $|y|<0.5$ for the granular source
for the RHIC collisions and the pion pseudorapidity satisfies
$|\eta|<0.8$ for the granular source for the LHC collisions,
respectively.  It can be seen that the HBT radii of the granular
sources agree well with the experimental data.  The wider
distributions of the transverse and longitudinal source points for
the LHC granular source (see Fig. \ref{Figdenstrz}) lead to the
larger transverse and longitudinal HBT radii than those for the RHIC
granular source.  In experimental HBT measurements, many effects,
such as the Coulomb interaction between the final particles,
particle missing-identification, source coherence, etc., can
influence the results of the $\lambda$
\cite{CYW94,UAW99,RMW00,MAL05}. Because we do not consider these
effects in our model, our $\lambda$ results are larger.

\begin{figure} [h]
\vspace*{5mm}
\includegraphics[scale=0.63]{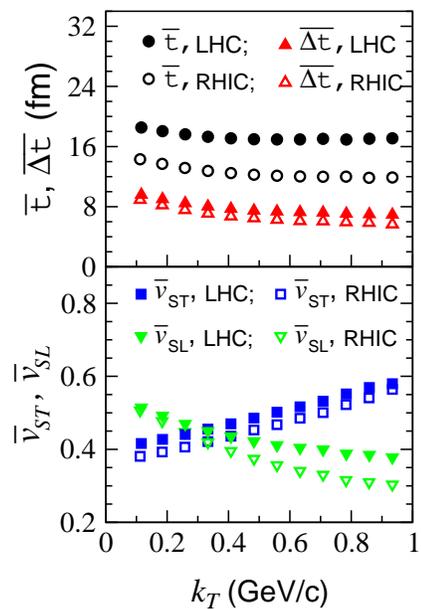}
\vspace*{1mm} \caption{(Color online) Upper panel: the average pion-emitting
time and relative pion-emitting time for the granular sources.  Lower panel:
the average source transverse and longitudinal velocities for the granular
sources.} \label{Ftv}
\end{figure}

The ratio of the transverse HBT radii $R_{\rm out}/R_{\rm side}$ is
related to the average relative emitting time of the two pions and
the velocity of the source transverse expansion, ${\overline {\Delta
t}}=\langle |t_1-t_2| \rangle$ and $v_{_{\rm ST}}$
\cite{UAW99,MAL05}. Although the larger breakup time $t_0$ for the
LHC source leads to a larger particle-emitting time, the limited
value of ${\overline {\Delta t}}$ for the granular sources
\cite{WNZ04,WNZ06,WNZ09,WNZ06a} and the large droplet transverse
velocity lead to the small $R_{\rm out}/R_{\rm side}$ results (see
Fig. \ref{Fighbtr} (e) and (e')).  In the upper panel of Fig.
\ref{Ftv}, we show the average pion-emitting time ${\overline
{\,t\,}}$ and the average relative pion-emitting time ${\overline
{\Delta t}}$ for the RHIC and LHC granular sources.  It can be seen
that the average pion-emitting time for the LHC source is much
larger than that for the RHIC source.  However, there is only small
difference between the average relative pion-emitting times for the
LHC and RHIC sources.  In the lower panel of Fig. \ref{Ftv}, we show
the average source transverse and longitudinal velocities as
functions of the transverse pion pair momentum $k_T$.  The
transverse velocity ${\overline v}_{_{\rm ST}}$ increases with
$k_T$, and the longitudinal velocity ${\overline v}_{_{\rm SL}}$
decreases with $k_T$.  For the collisions at the LHC energy the
granular source has larger transverse and longitudinal velocities.

In summary, we investigate the two-pion HBT interferometry in
ultrarelativistic heavy ion collisions in the granular source model
of QGP droplets.  The pion transverse momentum spectra and HBT radii
of the granular sources agree well with the RHIC $\sqrt{s_{NN}}=200$
GeV Au-Au and LHC $\sqrt{s_{NN}}=2.76$ TeV Pb-Pb most central
collisions.  Our results indicate that the granular source for the
collisions at the LHC energy may have the same initial droplet
temperature and velocity formula as those for the collisions at the
RHIC energy, but a larger initial system breakup time.  The larger
breakup time may lead to the larger longitudinal and transverse
distributions of the source points, and hence lead to the larger
longitudinal and transverse HBT radii.  However, the average
relative emitting time of the two pions is limited in the granular
source model.  The limited relative pion-emitting time and larger
droplet transverse expansion lead to the smaller ratio of the
transverse HBT radii, $R_{\rm out}/R_{\rm side}$.  In our granular
source model, there is correlations between the space-time
coordinates and source velocities.  The data of the particle
transverse momentum spectra and HBT radii for the collisions at the
RHIC and LHC energies give strong constraints for the model
parameters.  The consistent explain of the granular source model to
these experimental data will be helpful to understand the formation
and evolution of the pion-emitting sources in ultrarelativistic
heavy ion collisions.

\begin{acknowledgments}
This research was supported by the National Natural Science
Foundation of China under Contract No. 11075027.
\end{acknowledgments}

\end{document}